\documentstyle[12pt]{article}

\begin{document}
\pagestyle{empty}
\begin{flushright}
{CERN-TH/95-123 \\
 hep-lat/9505015}
\end{flushright}
\vspace*{5mm}
\begin{center}
{\bf INTERPOLATED LATTICE GAUGE FIELDS AND} \\[0.3cm]
{\bf CHIRAL FERMIONS IN THE SCHWINGER MODEL} \\
\vspace*{1.5cm}
{\bf I. Montvay$^1$}  \\
\vspace{0.3cm}
Theoretical Physics Division, CERN   \\
CH-1211 Geneva 23, Switzerland  \\
and  \\
Deutsches Elektronen-Synchrotron DESY, \\
Notkestr.\,85, D-22603 Hamburg, Germany  \\
\vspace*{3cm}
{\bf ABSTRACT} \\ \end{center}
\vspace*{3mm}
\noindent

 The effective action induced by fermions in the chiral Schwinger
 model with charges (3,4,5) is investigated.
 Pauli-Villars regularization is combined with momentum cut-off
 for the evaluation of the fermion determinants on continuum gauge
 fields interpolated between lattice points.
 The convergence and gauge variance are studied numerically on
 gauge configurations taken from quenched updating.
\vspace*{5cm}

\begin{flushleft}
CERN-TH/95-123 \\
May 1995

$^1$ e-mail address: montvay@surya20.cern.ch
\end{flushleft}
\vfill\eject

%%%%%%%%%%%%%%%%%%%%%%%%%%%%%%%%%%%%%%%%%%%%%%%%%
\newcommand{\be}{\begin{equation}}
\newcommand{\ee}{\end{equation}}
\newcommand{\half}{\frac{1}{2}}
\newcommand{\rar}{\rightarrow}
\newcommand{\lar}{\leftarrow}
\newcommand{\LCB}{\raisebox{-0.3ex}{\mbox{\LARGE$\left\{\right.$}}}
\newcommand{\RCB}{\raisebox{-0.3ex}{\mbox{\LARGE$\left.\right\}$}}}
%%%%%%%%%%%%%%%%%%%%%%%%%%%%%%%%%%%%%%%%%%%%%%%%%%%%%%

\setcounter{page}{1}
\pagestyle{plain}
\section{Introduction} \label{sec1}
 The non-perturbative definition of chiral gauge theories is a
 long-standing problem of quantum field theory on the lattice (for
 reviews see, for instance, \cite{SMIT}--\cite{CREUTZ}).
 Recently there have been several interesting new proposals trying to
 circumvent the obstacles represented by the Nielsen-Ninomiya theorem
 \cite{NINI} in different ways \cite{NARNEU}--\cite{BIEWIE}
 (for further references see these papers).

 In the present letter the idea put forward by 't Hooft \cite{THOOFT}
 is applied to the simple test case of the two-dimensional chiral
 Schwinger model.
 It is based on the interpolation of the lattice gauge field and
 the definition of the chiral fermion determinant on the obtained
 continuum gauge field, by exploiting the knowledge accumulated
 in continuum approaches (for a review see \cite{BALL}).
 Similar ways of defining chiral gauge theories on the lattice
 were discussed for some time \cite{FLUWYL,GOESCH}, and have been
 recently further developed in refs.~\cite{HSU,KRONFELD}.

 The aim of the present paper is to study numerically the
 definition of the effective action induced by chiral fermions
 on the interpolated lattice gauge field.
 As a first step, the two-dimensional massless chiral Schwinger model
 is considered here.
 This and related two-dimensional models are often used as a testing
 ground for chiral fermions (for recent examples see \cite{RECENT}).
 Since this model is well known and exacly soluble (in the extensive
 literature see, for instance, the papers in ref.~\cite{SCHW}), the
 questions are mainly oriented towards the qualitative behaviour of the
 calculation of the effective action along the line of
 refs.~\cite{THOOFT,HSU,KRONFELD}.
 The methods used will be such that they can be extended to four
 dimensions in a straightforward way.
 In the next section the interpolation of the U(1) gauge field is
 discussed.
 This is followed by a short discussion of some useful numerical
 algorithms.
 In section \ref{sec4} the convergence of the chiral fermion
 determinant is considered by removing the momentum cut-off.
 The variation with respect to gauge transformations is numerically
 investigated and discussed.
%%%%%%%%%%%%%%%%%%%%%%%%%%%%%%%%%%%%%%%%%%%%%%%%%%%%%%

\section{Gauge field interpolation} \label{sec2}
 The lattice gauge field is defined by the parallel transporters
 on the discrete links of the lattice, which is chosen in the
 present paper, for simplicity, to be hypercubic with periodic
 boundary conditions.
 To extend the gauge field connection into the meshes of the
 lattice is highly arbitrary.
 In order to reduce this arbitrariness some guiding principles must
 be respected, such as smoothness and some minimality
 principle which chooses among the different possibilities.
 Since in chiral gauge theories the anomaly plays an important
 r\^ole, one can also connect the gauge field interpolation
 to the geometrical definition of the topological charge
 \cite{LUESCH,GOKRSCHWI}.
 Both this interpolation and the piece-wise linear one
 minimizing the Euclidean action, which has been proposed in
 \cite{THOOFT}, have the important property that in momentum space
 the support of the Fourier transform is concentrated near the
 momenta allowed for the gauge field on the lattice.
 This has to be required as a condition for any reasonable
 interpolation: the momentum cut-off imposed on the gauge field by
 the lattice has to be approximately maintained by the interpolated
 gauge field too.

 Let us denote the U(1) gauge link variables on the lattice by
\be  \label{eq01}
U_{x\mu} = \exp( igA_{x\mu} ) \ .
\ee
 Here $x$ denotes lattice points: $x=(x_1,x_2)$ with integer $x_\mu$
 satisfying $0 \leq x_\mu \leq L_\mu-1,\; (\mu=1,2)$.
 The lattice extensions are denoted by $L_\mu$, $g$ is the bare
 gauge coupling and the number of lattice points will be denoted by
 $\Omega=L_1L_2$.
 Note that throughout this paper the lattice spacing of the lattice
 for the gauge field is set to be $a=1$.
 In other words, every dimensional quantity, as for instance the
 gauge field $A_{x\mu}$, is measured in lattice units of the gauge
 field lattice.
 The Fourier transformation to momentum space is defined, as usual, by
\be  \label{eq02}
\tilde{A}_{k\mu} \equiv \sum_x e^{-ik \cdot x - \frac{i}{2} k_\mu }
A_{x\mu} \ .
\ee
 The inverse relation is
\be  \label{eq03}
A_{x\mu} = \frac{1}{\Omega} \sum_k e^{ ik \cdot x } \tilde{A}_{k\mu}
\ ,
\ee
 where the sum is running, of course, on the points of the Brillouin
 zone
\be  \label{eq04}
k_\mu = \frac{2\pi}{L_\mu}\nu_\mu \ , \hspace{3em}
\nu_\mu = -{\rm int}(L_\mu/2),-{\rm int}(L_\mu/2)+1,
\; \ldots\; ,+{\rm int}(L_\mu/2) \ .
\ee

 The above discussed condition on the interpolation means that, if
 the Fourier transformation of the interpolated gauge field is
 performed on the continuous torus, the Fourier coefficients are
 approximately the same as in (\ref{eq02}).
 This suggests the introduction of eq.~(\ref{eq03}) as the definition
 of the interpolation by simply extending its validity to continuous
 $x$.
 This means that the interpolated gauge field on the continuum is,
 as a function of the continuous $x_c$:
\be  \label{eq05}
A_\mu(x_c) \equiv \sum_x D_\mu(x_c-x) A_{x\mu}
\ee
 with the {\em interpolation kernel}
\be  \label{eq06}
 D_\mu(x_c-x) \equiv \frac{1}{\Omega} \sum_k
e^{ik \cdot (x_c-x) - \frac{i}{2} k_\mu } \ .
\ee
 This is a very smooth interpolation indeed, since the result is
 infinitely many times differentiable (an entire function for complex
 $x_c$) satisfying, for integer $x$:
\be  \label{eq07}
A_\mu(x+\hat{\mu}/2) = A_{x\mu} \ .
\ee
 Here, as usual, $\hat{\mu}$ denotes the unit vector in direction
 $\mu$.

 Before going further, let us make a short technical remark.
 The Fourier coefficients in eq.~(\ref{eq02}) have the periodicity
 properties
\be  \label{eq08}
\tilde{A}_{k+2\pi\hat{\mu},\mu} = -\tilde{A}_{k,\mu} \ ,
\hspace{3em}
\tilde{A}_{k+2\pi\hat{\nu},\mu} = \tilde{A}_{k,\mu} \ ,
\ee
 where $\nu=3-\mu$.
 This allows us, for instance, to choose the symmetric interval for
 momenta in (\ref{eq04}).
 Of course, for the summation in the definition of $D_\mu$ in
 (\ref{eq06}) one has to specify the interval.
 In fact, for even $L_\mu$ it is advantageous to stick to an
 exactly symmetric definition by dividing the Fourier coefficient
 at $\nu_\mu=L_\mu/2$ into equal halfs at $\nu_\mu=L_\mu/2$ and
 $\nu_\mu=-L_\mu/2$.
 (For an odd $L_\mu$ the points in (\ref{eq04}) are automatically
 symmetric.)
 In the numerical study discussed in section \ref{sec4}
 this symmetric definition was always taken.

 It must be emphasized that the interpolation defined by
 eqs.~(\ref{eq05}) and (\ref{eq06}) is only one example among many
 others.
 It satisfies the important condition that the momentum cut-off for
 the gauge field be transferred from the lattice to the continuum.
 This is the only condition which will be exploited in what follows.
 This implies that the conclusions from the numerical study in
 section \ref{sec4} will be qualitatively valid also for the
 interpolations defined in refs. \cite{GOKRSCHWI,THOOFT}.

 In fact, from the point of view of gauge covariance the above
 definition satisfying (\ref{eq07}) is not optimal.
 Instead of it one can also require the alternative condition
\be  \label{eq09}
\int_x^{x+\hat{\mu}} dy A_\mu(y) = A_{x\mu} \ .
\ee
 This can be achieved, as one easily sees, by changing the definition
 of the continuation kernel in (\ref{eq06}) to
\be  \label{eq10}
 D_\mu(x_c-x) \equiv \frac{1}{\Omega} \sum_k \frac{k_\mu}{\hat{k}_\mu}
e^{ik \cdot (x_c-x) - \frac{i}{2} k_\mu } \ ,
\ee
 where $\hat{k}_\mu \equiv 2\sin(k_\mu/2)$.
 The consequence of eq.~(\ref{eq09}) is gauge covariance.
 Performing the gauge transformation on the lattice by the U(1)
 elements $\Lambda_x = \exp(i\alpha_x)$ and continuing $\alpha_x$
 to a function $\alpha(x_c)$ on the continuum in such a way that,
 for integer $x_c=x$,
\be \label{eq11}
\alpha(x) = \alpha_x \ ,
\ee
 we obtain the relation for the gauge-transformed links
\be \label{eq12}
U^{(\Lambda)}_{x\mu} = \Lambda_{x+\hat{\mu}}^{-1}U_{x\mu}\Lambda_x
= \exp\left\{ ig \int_x^{x+\hat{\mu}} dy A^{(\Lambda)}_\mu(y)
\right\} \ .
\ee
 Here $A^{(\Lambda)}_\mu(x)$ is the gauge-transformed continuum
 field
\be \label{eq13}
A^{(\Lambda)}_\mu(x) = A_\mu(x) - g^{-1} \partial_\mu \alpha(x) \ .
\ee
 A convenient interpolation satisfying eq.~(\ref{eq11}) is given by
\be  \label{eq14}
\alpha(x_c) \equiv \sum_x D(x_c-x) \alpha_x
\ee
 with the continuation kernel for scalar fields, analogous to
 $D_\mu$ in (\ref{eq06}) or (\ref{eq10}),
\be  \label{eq15}
 D(x_c-x) \equiv \frac{1}{\Omega} \sum_k e^{ik \cdot (x_c-x)} \ .
\ee

 The gauge transformations become relevant if the effective action
 is not exactly gauge invariant (see the discussion in section
 \ref{sec4}).
 The choice of a gauge can be made on the basis of the minimality
 principle.
 This leads to the {\em Landau gauge}, which can be defined on the
 lattice by requiring
\be  \label{eq16}
f_2[U] \equiv \sum_x \sum_{\mu=1}^2 A_{x\mu}^2
\ee
 to be minimal with respect to gauge transformations.
 For U(1) gauge fields there are efficient algorithms to find this
 absolute minimum, for instance the one discussed in \cite{GAUFIX},
 which will be used in the present paper.

 The advantage of the gauge-field interpolation given by
 eqs.~(\ref{eq05}), (\ref{eq10}) is its simplicity and the direct
 relation to momentum space, which will be useful for the evaluation
 of the determinant in momentum basis.
 Concerning topological charge, it defines a continuous gauge field
 on the torus which has a total classical topological charge zero.
 On large volumes this is not a serious constraint, because the parts
 of the volume can have any topological charge.
 In fact, the interpolated gauge field can be used to define a
 topological charge density operator with the appropriate
 renormalization procedure for composite operators.
 (See \cite{ACDGV} and references therein.)

%%%%%%%%%%%%%%%%%%%%%%%%%%%%%%%%%%%%%%%%%%%%%%%%%%%%%%

\section{Computing the determinant} \label{sec3}
 The Euclidean action for massless chiral fermions in the U(1)
 background gauge field $A_\mu(x)$ is given in the continuum by
\be  \label{eq17}
S = \int d^2x \left\{
\overline{\psi}(x) \gamma_\mu\partial_\mu \psi(x)
- igA_\mu(x)\overline{\psi}(x) \gamma_\mu
\left( P_R Q_R+P_L Q_L \right) \psi(x) \right\} \ .
\ee
 Here $P_R \equiv (1+\gamma_3)/2$ and $P_L \equiv (1-\gamma_3)/2$ are
 the chiral projectors for right-handed and left-handed fermions,
 respectively.
 We shall use the Pauli matrices for the $\gamma$-matrices in two
 dimensions: $\gamma_\mu \equiv \sigma_\mu,\; (\mu=1,2,3)$.
 $Q_R$ and $Q_L$ are the charges of the chiral fermion components.
 Note that in (\ref{eq17}) we implicitly adopt the ``doubling trick''
 \cite{ALVGIN,ALVDEL}: even if one of the charges $Q_{R,L}$ is zero,
 we use a Dirac fermion field.
 In this way the chiral fermion determinant is always a
 {\em determinant} indeed.

 The fermion matrix in momentum space corresponding to
 eq.~(\ref{eq17}) is
\be  \label{eq18}
M_{k_2k_1}^{Q_R,Q_L} = \Omega \delta_{k_2k_1} i\gamma \cdot k_1
-ig\gamma_\mu \left( P_R Q_R+P_L Q_L \right)
\tilde{A}_{k_2-k_1,\mu} \ .
\ee
 After multiplication by the fermion propagator we obtain
\be  \label{eq19}
N_{k_2k_1}^{Q_R,Q_L} \equiv M_{k_2k_1}^{Q_R,Q_L}
\frac{(-i\gamma \cdot k_1)}{\Omega k_1^2}
\equiv \delta_{k_2k_1} - K_{k_2k_1}^{Q_R,Q_L} \ .
\ee
 This has the following matrix elements in spinor indices:
$$
N_{k_2k_1}^{Q_R,Q_L}(1,1) =
\delta_{k_2k_1} - Q_L (a_{k_2-k_1,1}-ia_{k_2-k_1,2})
(k_{1,1}+ik_{1,2})/k_1^2  \ ,
$$
$$
N_{k_2k_1}^{Q_R,Q_L}(1,2) = 0  \ ,
$$
$$
N_{k_2k_1}^{Q_R,Q_L}(2,1) = 0  \ ,
$$
\be  \label{eq20}
N_{k_2k_1}^{Q_R,Q_L}(2,2) =
\delta_{k_2k_1} - Q_R (a_{k_2-k_1,1}+ia_{k_2-k_1,2})
(k_{1,1}-ik_{1,2})/k_1^2 \ .
\ee
 Here the explicit representation of the $\gamma$-matrices and the
 short notation $a \equiv g\tilde{A}/\Omega$ is used.
 For a massive vector-like fermion with mass $m$ and $Q\equiv Q_R=Q_L$,
 which will be used for Pauli-Villars fields, the matrix elements
 corresponding to (\ref{eq20}) are:
$$
N_{k_2k_1}^{Q(m)}(1,1) =
\delta_{k_2k_1} - Q (a_{k_2-k_1,1}-ia_{k_2-k_1,2})
(k_{1,1}+ik_{1,2})/(m^2+k_1^2)  \ ,
$$
$$
N_{k_2k_1}^{Q(m)}(1,2) =
-imQ (a_{k_2-k_1,1}-ia_{k_2-k_1,2})/(m^2+k_1^2)  \ ,
$$
$$
N_{k_2k_1}^{Q(m)}(2,1) =
-imQ (a_{k_2-k_1,1}+ia_{k_2-k_1,2})/(m^2+k_1^2)  \ ,
$$
\be  \label{eq21}
N_{k_2k_1}^{Q(m)}(2,2) =
\delta_{k_2k_1} - Q (a_{k_2-k_1,1}+ia_{k_2-k_1,2})
(k_{1,1}-ik_{1,2})/(m^2+k_1^2) \ .
\ee

 For the computation of the determinants of these matrices in
 momentum basis an appropriate algorithm is the LU (lower-upper
 triangular) decomposition (see, for instance, \cite{YOUGRE}).
 It turned out to be both robust and sufficiently fast on the
 lattices considered.
 It can also be used for the computation of the full inverse
 matrix, and the algorithm can be organized in such a way that
 the matrix has to be stored only once.
 Of course, storing these large matrices even only once is the main
 limiting factor of the computation.
 For very large matrices also the time requirement is growing
 dangerously: it behaves as the third power of the matrix
 extension.

 In order to extend the range of feasible lattice sizes one can
 exploit some additional iterative procedures.
 Before describing them let us discuss the momentum cut-off
 scheme used.
 Since, according to the previous section, the Fourier components
 of the gauge field are constrained to the points (\ref{eq04}) of
 the Brillouin zone belonging to the gauge field lattice, it
 is natural to use a momentum cut-off for the calculation of the
 determinants of the infinite-dimensional matrices in
 (\ref{eq19})--(\ref{eq21}).
 One can imagine to make the lattice finer for the fermions by adding
 more points to the gauge field lattice.
 In this case, however, the periodicity in momentum components is
 maintained, which introduces some non-zero elements also near the
 upper right and lower left corner, besides the ones near the main
 diagonal, for which $a_{k_2-k_1,\mu} \ne 0$.
 This makes the effect of the cut-off stronger, therefore it is
 more advantageous to abandon periodicity and drop the extra non-zero
 elements.
 In this way, for momentum cut-offs much larger than $\pi$ (in gauge
 field lattice units), the matrix has a band structure.

 As a consequence of this band structure, one can effectively apply
 the iterative algorithm previously used for the numerical hopping
 parameter expansion in QCD \cite{HOPPING}.
 For this one determines the traces of the powers of the {\em hopping
 matrix} $K$ (here in momentum space).
 Having these traces one can use either the usual infinite expansion
\be  \label{eq22}
\det(1-K) = \exp \left\{ -\sum_{j=1}^\infty
\frac{{\rm Tr\,}K^j}{j} \right\} \ ,
\ee
 or the finite {\em polymer representation} \cite{KERLER}
\be  \label{eq23}
\det(1-K) = 1 + \sum_{\nu=1}^n \sum_{r=1}^\nu \frac{(-1)^r}{r!}
\sum_{\rho_1=1}^{\nu-r+1} \ldots \sum_{\rho_r=1}^{\nu-r+1}
\delta_{\nu,\rho_1+\ldots+\rho_r}
\frac{{\rm Tr\,}K^{\rho_1}}{\rho_1}
\frac{{\rm Tr\,}K^{\rho_2}}{\rho_2} \ldots
\frac{{\rm Tr\,}K^{\rho_r}}{\rho_r}  \ .
\ee
 This latter is always convergent because it is a sum of a finite
 number of terms.
 In practical calculations one can go without any problems to
 $j_{\rm max} \simeq 100$ in eq.~(\ref{eq22}) or to
 $n_{\rm max} \simeq 40$ in eq.~(\ref{eq23}).
 As a consequence of the band structure of the matrices $K$ in
 (\ref{eq19})--(\ref{eq21}), the storing of the full matrices is
 not necessary, and the computational load is growing as the second
 power of the matrix extensions times  $j_{\rm max}^2$ or
 $n_{\rm max}^2$.

 Inspection of the matrices in (\ref{eq19})--(\ref{eq21}) shows that
 only the traces of even powers are non-zero.
 One can also easily see that
\be  \label{eq24}
\left[ {\rm Tr}\left( K^{Q_R,Q_L} \right)^{\rho} \right]^* =
{\rm Tr}\left( K^{Q_L,Q_R} \right)^{\rho} \ .
\ee
 This corresponds to the relation
\be  \label{eq25}
\left[ \det N^{Q_R,Q_L} \right]^* = \det N^{Q_L,Q_R} \ .
\ee
 Therefore the determinant of the vector-like fermion
 $\det N^{Q(m)}$ is real.
 (One can also easily prove that it is non-negative.)

 Concerning the practical convergence of the {\em trace expansions} in
 (\ref{eq22}) and/or (\ref{eq23}) in the chiral Schwinger
 model with charges (3,4,5) the experience is negative.
 Typically neither of them converges, because in the calculated
 range the contributions rapidly increase.
 This is mainly the consequence of the large value of the charges
 (see next section).
 One can, however, easily save their advantages by calculating the
 determinant and inverse of the matrices $N$ truncated
 to a smaller sublattice (typically of the same size as the lattice
 for the gauge field), and then use
\be  \label{eq26}
\det N = \det N_{\rm small} \det[N_{\rm small}^{-1}(1-K)] =
\det N_{\rm small} \det(1-K_{\rm new}) \ ,
\ee
 with
\be  \label{eq27}
K_{\rm new} \equiv 1 - N_{\rm small}^{-1}(1-K)  \ .
\ee
 It turns out that the expansions in the traces of the powers of
 $K_{\rm new}$ converge rapidly.
 The omitted higher-order terms could always be kept smaller than
 $10^{-8}$, relative to the result.

%%%%%%%%%%%%%%%%%%%%%%%%%%%%%%%%%%%%%%%%%%%%%%%%%%%%%%

\section{Convergence and gauge variance} \label{sec4}
 The commonly considered example of an anomaly-free chiral Schwinger
 model has fermion charges $(Q_R=3,Q_L=0)$, $(Q_R=4,Q_L=0)$ and
 $(Q_R=0,Q_L=5)$.
 (The anomaly is cancelled if the sum of squared charges of the
 right-handed fermions is equal to those of the left-handed ones.)
 In order to regulate the ultraviolet divergence of the two-point
 function in a gauge-invariant way, one can introduce a Pauli-Villars
 vector-like fermion field with charge $Q=5$ \cite{BAFRYA}.
 In this way the effective action $S_{\rm eff}$ induced by the
 fermions is given by
\be  \label{eq28}
\exp\{ -S_{\rm eff}(M_{PV}) \} \equiv E_{\rm eff}(M_{PV}) =
\det N^{3,0} \det N^{4,0} \det N^{0,5}/\det N^{5(M_{PV})}  \  .
\ee
 Here $M_{PV}$ is the mass of the Pauli-Villars regulator field in units
 of the gauge field lattice.
 As argued in ref.~\cite{KRONFELD}, in the continuum limit $M_{PV}$
 should be kept finite, for instance of order 1.
 This is necessary in order to maintain the possibility of a simple
 renormalization.

 Some insight into the behaviour of the effective action defined by
 eq.~(\ref{eq28}) can be obtained by numerically evaluating the
 determinants on some typical gauge configurations taken from Monte
 Carlo updating.
 For this I took quenched updating by the usual compact U(1) gauge
 field Wilson action.
 The gauge coupling has been fixed by $\beta \equiv g^{-2} = 8$, which
 is a quite strong coupling for these fermions.
 Namely, the interaction strength is given by $Qg$, therefore
 weak couplings are beyond $\beta \simeq 25$.
 The gauge configurations were transformed to Landau gauge by the
 algorithm described in \cite{GAUFIX}.
 The gauge field lattices were either $4 \otimes 4$ or
 $10 \otimes 10$.
 The lattices defining the momentum cut-off for the evaluation of
 the determinants always had an odd number of points (see discussion
 after eq.~(\ref{eq08})), and they went up to $61 \otimes 61$.
 This means that the momentum cut-offs went up roughly to $15\pi$
 (in units of the gauge field lattice).
 For the Pauli-Villars mass values between $M_{PV}=\half$ and
 $M_{PV}=\pi$ were tried.

\begin{figure}[h]
\vspace{10.0cm}
\includegraphics{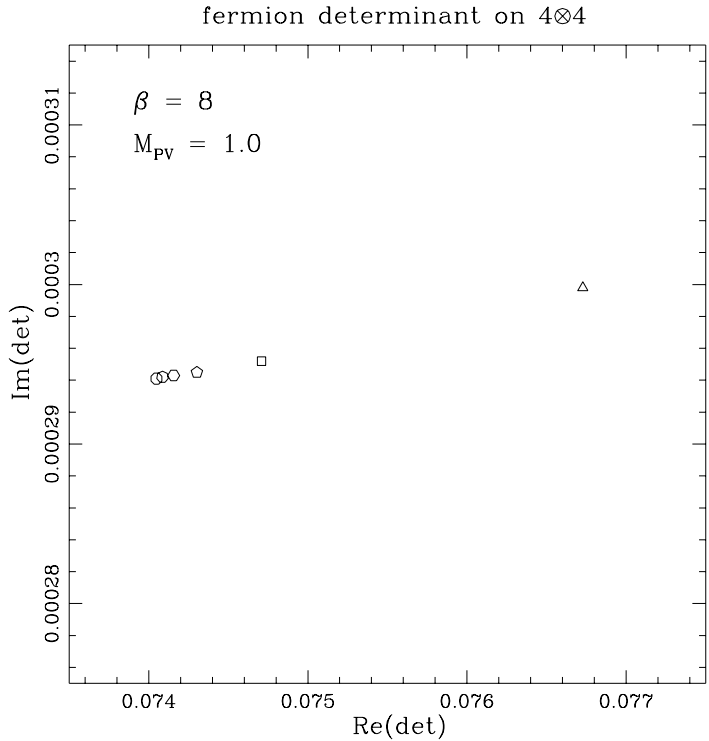}
\begin{center}
\parbox{16cm}{\caption{ \label{fig1}
 The values of $E_{\rm eff}(1)$ in the complex plane on a
 $4 \otimes 4$ gauge field with momentum cut-off on $11 \otimes 11$
 (triangle), $21 \otimes 21$ (quadrangle), etc., up to $61 \otimes 61$
 (eight-angle).
}}
\end{center}
\end{figure}

 A first important question is how fast the infinite cut-off limit
 is reached by the effective action.
 It turned out that for every considered gauge configuration, taken
 randomly from the updating and transformed to Landau gauge, a good
 convergence could be achieved with the above cut-offs, provided
 that $M_{PV}$ was not too large.
 For an illustration on $4 \otimes 4$ lattice see fig.~\ref{fig1}.
 The numerical results for nine configurations on $10 \otimes 10$ are
 shown in table \ref{tab1}.

 It is interesting to investigate the gauge dependence of the
 determinants.
 The gauge transformation of the infinite-dimensional fermion matrix
 $M$ in (\ref{eq18}) is given by $\Lambda^\dagger M \Lambda$, where
 in momentum space
\be  \label{eq29}
\Lambda_{k_2k_1} = \frac{1}{\Omega} \int d^2x
e^{ ix \cdot (k_1-k_2) + i\alpha(x) } =
\delta_{k_2k_1} + \frac{i}{\Omega}\tilde{\alpha}_{k_2-k_1}
+ \frac{i^2}{\Omega^2} \sum_k \tilde{\alpha}_k
\tilde{\alpha}_{k_2-k_1-k} + \ldots  \ ,
\ee
 with $\tilde{\alpha}_k$ denoting the Fourier components of
 $\alpha(x)$.
 This infinite-dimensional unitary matrix is truncated by the
 momentum cut-off.
 Therefore, gauge invariance of the fermion determinant is lost even
 for vector-like fermions, which were gauge-invariant without
 truncation.
 The chiral fermion determinants remain gauge non-in\-vari\-ant
 also for infinite momentum cut-off.
%%%%%%%%%%%%%%%%%%%%%%%%%%%%%%%%%%%%%%%%%%%%%%%%%%%%%%%%%%%%%%%%%%
%%%%%%%%%%%%%%%%%%%%%%%%%%%%%%%%%%%%%%%%%%%%%%%%%%%%%%%%%%%%%%%%%%

%%%%%%%%%%%%%%%%%%%%%%%%%%%%%%%%%%%%%%%%%%%%%%%%%%%%%%%%%%%%%%%%%%
\begin{figure}[t]
\vspace{10.0cm}
\includegraphics{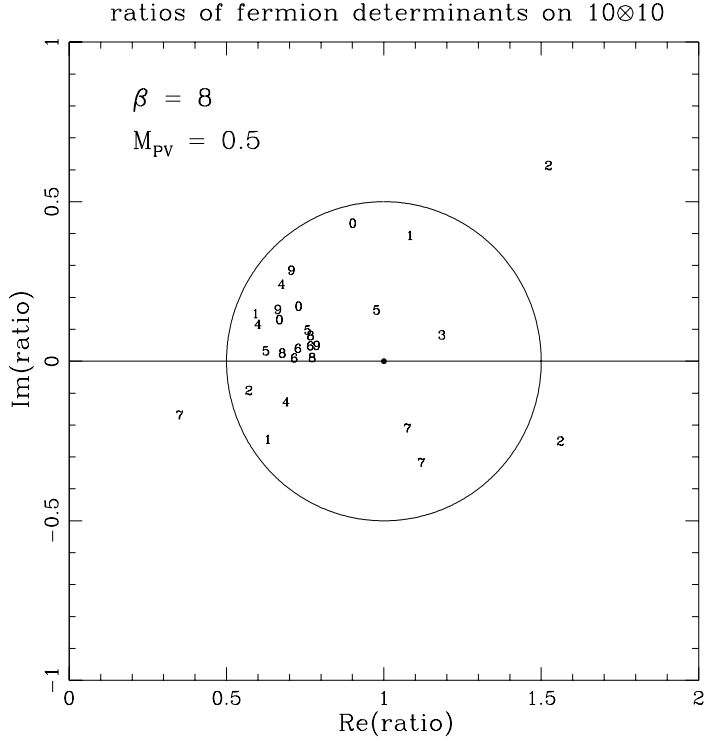}
\begin{center}
\parbox{16cm}{\caption{ \label{fig2}
 The residual gauge variance of the effective action shown by the
 (complex) ratios of $E_{\rm eff}(\half)$ atfer and before gauge
 transformation.
 The numbers are labeling different configurations.
 The circle with radius $\half$ around the point (1,0) is drawn to
 guide the eyes.
}}
\end{center}
\end{figure}
%%%%%%%%%%%%%%%%%%%%%%%%%%%%%%%%%%%%%%%%%%%%%%%%%%%%%%%%%%%%%%%%%%
 The variation with gauge transformations is displayed in figure
 \ref{fig2} for ten $\beta=8$ gauge configurations on $10 \otimes 10$
 lattice.
 The configurations were first transformed to Landau gauge and
 then random gauge transformations were performed with parameters
 satisfying on the lattice points $-\pi/20 < \alpha_x < \pi/20$.
 The gauge transformations were interpolated in the continuum as
 described in section \ref{sec2}.
 The determinants were calculated on $31 \otimes 31$ (momentum
 cut-off $=3.1\pi$).
 As is shown in the figure, the gauge variation is not very strong.
 In fact, both numerator and denominator of $E_{\rm eff}(\half)$
 in (\ref{eq28}) always change by 4--5 orders of magnitude, but the
 ratio remains close to 1.
 Performing random gauge transformations with larger magnitude
 (up to  $-\pi < \alpha_x < \pi$) shows an ever-increasing change
 in numerators and denominators, such that it becomes difficult to
 keep the numbers in the computer, but the overwhelming
 part of the variation is cancelled in the ratio.
 The cancellation can be further improved by taking more
 Pauli-Villars fields with appropriately chosen larger masses.

 There are two possibilities for dealing with this residual gauge
 variation of the effective action.
 First, one can try to tolerate it, keeping the momentum cut-off
 finite in gauge field lattice units.
 Second, more radically, one can enforce exact gauge invariance by
 defining the effective action to be equal to its value in Landau
 gauge along the whole gauge transformation orbit.
 The hope is that at the end, in the continuum limit, both these
 procedures lead to the same well defined theory.

 The gauge-field interpolation combined with momentum cut-off for
 the evaluation of the Pauli-Villars regulated determinants seems
 to work reasonably well in the (3,4,5) chiral Schwing\-er model.
 It can be expected that the effective action defined in this way
 leads to a well defined continuum limit.
 Of course, momentum cut-off is not the only possibility.
 Examples of other possibilities are, for instance, to take a finer
 lattice (in coordinate space) for fermions and to use the formalism
 of ref.~\cite{ALVDEL} for the imaginary part of the effective action
 as suggested in \cite{GOESCH}, or to take on the finer fermion
 lattice the SLAC derivative, as proposed by ref.~\cite{SLAVNOV}.
 One has to see which one of these (or some other) approaches has the
 most conceptual and practical advantages.

%%%%%%%%%%%%%%%%%%%%%%%%%%%%%%%%%%%%%%%%%%%%%%%%%%%%%%%%%%%%%%%%%%
\vspace{6em}
\begin{table}[h]
\begin{center}
\parbox{14cm}{\caption{ \label{tab1}
 The values of the determinants on $10 \otimes 10$ gauge field
 configurations with momentum cut-off $5.1\pi$.
 The first line for a given configuration label is the value on
 the ``small'' subspace $\det N_{\rm small}$, the second line
 the correction factor obtained by trace expansion.
 For the configurations above the double line $N_{\rm small}$ is with
 momentum cut-off $2.1\pi$, below it with $3.1\pi$.
 The complex numbers are given by pairs in parentheses.
}}
\end{center}
\begin{center}
\begin{tabular}{|c|c|c|c|c|}
\hline
  & $\det N^{3,0}$ & $\det N^{4,0}$ & $\det N^{0,5}$ &
$\det N^{5(\half)}$  \\
\hline
1 & $(2.1481,-0.8303)$  & $(0.3880,0.5783)$  &
  $(-68.1036,-40.6948)$ & $3.5545 \times 10^{10}$       \\
  & $(1.0216,-0.0055)$  & $(1.0364,-0.0288)$ &
    $(1.0541,0.0308)$   & 1.3258                        \\
\hline
2 & $(0.4173,-0.2723)$  & $(0.2104,-0.2211)$  &
    $(-0.0557,-0.1826)$ & $5.5351 \times 10^{8}$        \\
  & $(1.0374,-0.0053)$  & $(1.0672,-0.0148)$ &
    $(1.1058,0.0367)$   & 1.4123                        \\
\hline
3 & $(0.0987,0.0139)$   & $(0.0169,0.0081)$  &
    $(0.0020,-0.0020)$  & $5.6234 \times 10^{4}$        \\
  & $(1.0482,-0.0000)$  & $(1.0892,-0.0040)$ &
    $(1.1457,0.0155)$   & 1.4466                        \\
\hline
4 & $(0.6872,0.2887)$   & $(1.2979,0.1028)$  &
    $(1.2729,0.0435)$   & $9.3387 \times 10^{4}$        \\
  & $(1.0303,0.0009)$   & $(1.0561,0.0052)$  &
    $(1.0929,-0.0166)$  & 1.3091                        \\
\hline
5 & $(0.3724,0.0593)$   & $(0.1504,-0.0241)$  &
    $(0.0073,0.0931)$   & $1.3400 \times 10^{4}$        \\
  & $(1.0268,-0.0073)$  & $(1.0458,-0.0117)$  &
    $(1.0673,0.0132)$   & 1.2518                        \\
\hline
6 & $(0.2303,0.0165)$   & $(0.0764,0.0111)$  &
    $(0.0156,-0.0025)$  & $2.4780 \times 10^{1}$        \\
  & $(1.0271,-0.0008)$  & $(1.0482,-0.0004)$ &
    $(1.0751,-0.0007)$  & 1.2412                        \\
\hline
7 & $(1.3350,-0.3054)$  & $(2.0516,0.3677)$  &
   $(-1.1129,-17.2413)$ & $3.6316 \times 10^{10}$       \\
  & $(1.0474,0.0084)$   & $(1.0884,0.0262)$  &
    $(1.1481,-0.0658)$  & 1.5387                        \\
\hline\hline
8 & $(0.00956,-0.00109)$    & $(0.000327,-0.000086)$  &
    $(0.0000037,0.0000019)$ & $5.920 \times 10^{8}$     \\
  & $(1.0402,-0.0048)$      & $(1.0322,-0.6017)$  &
    $(0.8859,-0.1220)$      & 1.3736                    \\
\hline
9 &  $(0.00945,0.00077)$     & $(0.000336,0.000054)$  &
    $(0.0000042,-0.0000024)$ & $1.600 \times 10^{8}$    \\
  & $(1.0404,-0.0009)$       & $(1.0257,-0.0070)$  &
    $(0.7122,0.2096)$        & 1.3379                   \\
\hline
\end{tabular}
\end{center}
\end{table}
%%%%%%%%%%%%%%%%%%%%%%%%%%%%%%%%%%%%%%%%%%%%%%%%%%%%%%%%%%%%%%%%%%

%%%%%%%%%%%%%%%%%%%%%%%%%%%%%%%%%%%%%%%%%%%%%%%%%%%%%%%%%%%%%%%%%%

%%%%%%%%%%%%%%%%%%%%%%%%%%%%%%%%%%%%%%%%%%%%%%%%%%%%%%%%%%%%%%%%%%

\end{document}